%% file: J1713event_manuscript_arXiv.tex
\documentclass[twocolumn]{aastex701}

\usepackage{xcolor}

\def\be{\begin{equation}}
\def\ee{\end{equation}}

\begin{document}

\title{Mitigating the Timing Impact of Anomalous Pulse Profile Shape Variability in PSR~J1713+0747 with Gaussian Component Modeling}

\input{authorsJ1713event_arXiv}

\begin{abstract}
The North American Nanohertz Observatory for Gravitational Waves (NANOGrav) achieves sub-microsecond timing precision for several millisecond pulsars in its pulsar timing array (PTA) with the objective of detecting and characterizing nanohertz gravitational waves. PSR~J1713+0747 is one of the most precisely timed pulsars in the array, achieving sub-microsecond timing precision. However, in April 2021, PSR~J1713+0747 underwent a sudden and unusual change in pulse shape that disrupted its timing stability. As PSR~J1713+0747 is a key contributor to PTA sensitivity, variations in its pulse profile significantly affect the array's sensitivity to nanohertz gravitational waves. We apply frequency-dependent Gaussian component models to decompose the pulse profile and track the evolution of individual components through the event. This component-level method maintains phase-connected timing across the shape-change event. At L-band, the recovered TOAs have a median uncertainty of $\sim0.47~\mu$s compared to $\sim0.69~\mu$s for standard template matching. At 820~MHz, where profile evolution is stronger, the recovered TOAs have a median uncertainty of $\sim1.63~\mu$s compared to $\sim0.67~\mu$s for standard template matching. The recovered TOAs achieve timing uncertainties comparable to conventional template matching while allowing data affected by profile variability to be retained in PTA gravitational-wave analyses. These results represent an initial step toward profile-domain timing methods capable of accounting for pulse-profile evolution while reducing the need for additional timing model parameters.
\end{abstract}

%\keywords{{Pulsars: individual: {PSR~J1713+0747}}}

\section{Introduction} \label{intro}

PSR~J1713+0747 is observed by multiple pulsar timing arrays (PTAs) and plays an integral role in the characterization of nanohertz gravitational waves.
In 2023, the North American Nanohertz Observatory for Gravitational Waves (NANOGrav) Collaboration, together with members of International Pulsar Timing Array (IPTA) and the Chinese Pulsar Timing Array (CPTA) published evidence for the nanohertz gravitational-wave background \citep{NANOGrav:2023gor, EPTA:2023fyk, Reardon:2023gzh, Xu:2023wog}.
The sensitivity of PTAs to nanohertz gravitational waves increases the longer the timing baselines of the pulsars are \citep{Siemens:2013zla}.
Maximizing the span of timing data available for gravitational-wave background analysis is essential for characterization. 

\begin{figure}
    \centering
    \includegraphics[width=1.0\linewidth]{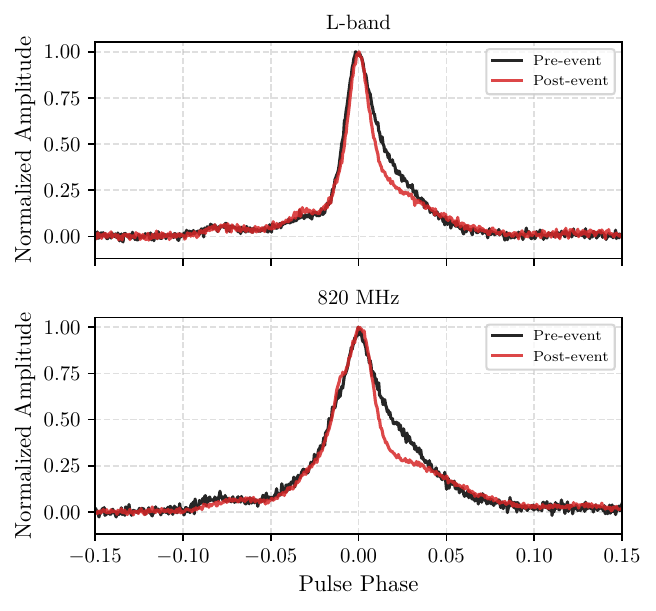}
    \caption{Frequency averaged pulse profile for single day before (black) and after (red) the shape change event (MJD 59320-59321). The top panel shows the L-Band profile, and the bottom panel shows the 820~MHz profile.
}
    \label{fig:profiles}
\end{figure}

Pulse profile stability is a key assumption underlying high-precision pulsar timing methods. 
The millisecond pulsar (MSP) J1713+0747 experienced an abrupt and significant change in its integrated pulse profile between April 16 and 17, 2021 (Modified Julian Date (MJD) 59320-59321; see Figure \ref{fig:profiles}; \citep{Xu2021ATel14642, Jennings:2024stq}). 
Although the origin of the shape change is unknown, it is likely magnetospheric in nature as interstellar medium (ISM) propagation effects do not sufficiently explain the complexity of the profile morphology \citep{Jennings:2024stq}. 
Recent broadband polarimetric observations further support a magnetospheric origin, showing frequency-dependent and polarization-dependent profile evolution, while the pulse profile remains distinct from its pre-event state more than three years after the event \citep{Mandow:2025vju}.

An ideal pulse profile template for timing analysis is narrow in phase and stable in time, which ensures that the pulse times of arrival (TOAs) reflect the rotational behavior of the pulsar and known astrophysical contributions, rather than instrumental effects.
Temporal evolution in the integrated profile shape of a pulsar introduces systematic biases in template matching, leading to reduced precision in pulse TOA measurements \citep{Shannon:2016icc, Nathan:2023qwn, Jennings:2024stq}.
Although PSR~J1713+0747 is one of the brightest and most precisely timed pulsars by NANOGrav \citep{Demorest:2012bv, Zhu:2015mdo}, subtle shape changes can produce measurable deviations in timing residuals thereby impacting the sensitivity of PTA experiments.

Long-term observations of PSR~J1713+0747 reveal two previous chromatic, or frequency-dependent, events causing perturbations in pulse times of arrival within the first 15 years of NANOGrav data \citep{Demorest:2012bv, Lam:2017duo}.
These events initially appeared consistent with frequency-dependent dispersion variations, although subsequent reanalysis of the second event identified subtle pulse-shape variations associated with the timing perturbation \citep{Lin:2021vzj}.
These findings emphasize that profile variability is not unique to the 2021 event and highlights the need for timing methods that remain robust to such changes. 
Pulse-profile shape variations have previously been observed in other PTA millisecond pulsars, including PSR~J1643$-$1224, where a transient profile disturbance was interpreted as a magnetospheric event \citep{Shannon:2016icc}.
As PTA datasets continue to expand and more pulsars are added to the IPTA, similar profile–shape change events may be encountered with increasing frequency.
It is imperative to understand and mitigate the timing impact of such profile-shape changes.

\begin{figure}
    \centering
    \includegraphics[width=1.0\linewidth]{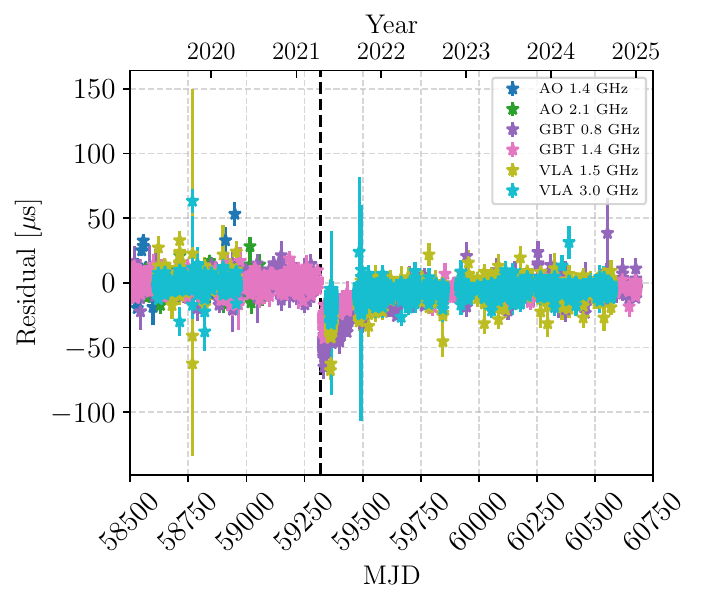}
    \caption{Preliminary timing residuals for PSR~J1713+0747 from the NANOGrav 20-year dataset, including observations from the Arecibo Observatory (AO), the Green Bank Telescope (GBT), and the Very Large Array (VLA), shown over the interval MJD 58500–60750 surrounding the profile-shape change event that began at MJD~59320 (black dashed line). The TOAs are generated using a constant-in-time standard profile, and the timing model used to fit them is derived from the preliminary NANOGrav 20-year dataset. The event produces frequency-dependent deviations in the timing residuals, demonstrating that pulse-profile evolution introduces systematic offsets in the measured TOAs. Residuals remain perturbed for several years following the onset of the event.
}
    \label{fig:toas}
\end{figure}

The April 2021 event manifests as a profile deviation, therefore the estimated timing residuals deviate rapidly corresponding to the mismatch between the evolving pulse shape and the static template used in standard timing profiles (Figure~\ref{fig:toas}). 
Without a method that is able to track and model these variations, the affected TOAs are biased, reducing the available data and shortening the effective timing baseline used for nanohertz gravitational-wave background detection.

\begin{figure}
    \centering
    \includegraphics[width=1.0\linewidth]{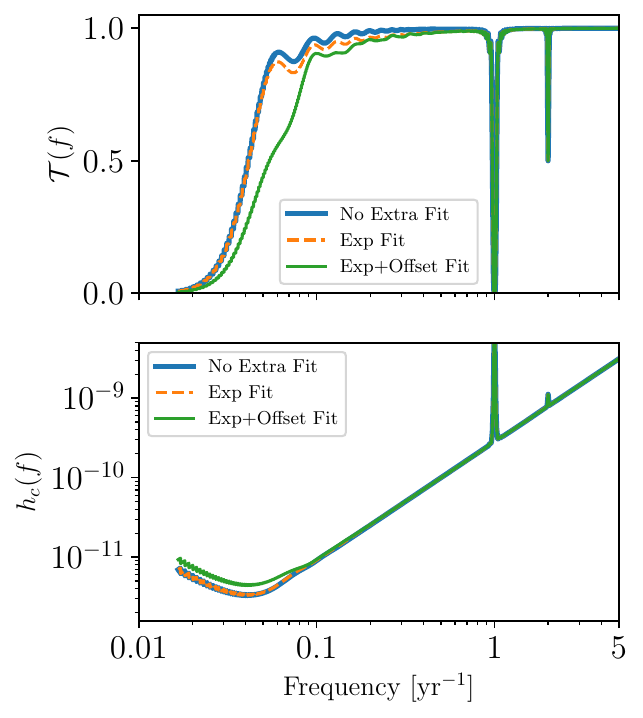}
    \caption{\textit{Top Panel:} Transmission function, $\mathcal{T}(f)$, for a cartoon 30-year PTA dataset. The three curves correspond to different treatments of the profile-shape change event: (i) no additional event modeling, corresponding to appropriate corrections of the timing perturbation; (ii) an exponential recovery with a fixed decay timescale of 156 days occurring 16.2 years after the start of the dataset \citep{Jennings:2024stq}; and (iii) the same exponential recovery model with an additional timing offset following the event. All models include the standard pulsar spin and astrometric parameters. \textit{Bottom Panel:} Corresponding sky-averaged characteristic strain sensitivity curves, $h_c(f)$, for the three timing models shown in the top panel. A white-noise level of 100~ns is assumed throughout the 30-year dataset. Frequency-dependent effects, including chromatic timing perturbations and frequency-dependent event modeling, are not included in this illustrative calculation.
}

    \label{fig:transmission}
\end{figure}

Fitting any functional form to the timing model, such as an exponential decay shape or an exponential decay shape plus an offset that accounts for permanent profile differences, reduces any gravitational wave power in the residuals. Figure~\ref{fig:transmission} shows transmission functions, $\mathcal{T}(f)$, representing the reduction in power at each Fourier frequency due to simplified timing models \citep{1984JApA....5..369B}, assuming a total observing timespan of 30 years, effectively demonstrating what happens to a future dataset if we require additional timing model parameters or can develop an upstream profile-based model which can correct the TOAs appropriately. 
We also show the corresponding characteristic strain sensitivity curves \citep{2019PhRvD.100j4028H} assuming a constant 100~ns of white noise. 
All three timing models have spin and astrometric terms fit for while the second includes an exponential decay term and the third includes exponential decay plus offset terms. 
We have fixed the decay constant to 156 days starting 16.2 years after the start of the observations \citep{Jennings:2024stq}. 
We see that the exponential fit reduces power, but only slightly, and that fitting the unknown offset is worse; this is a known effect \citep[see the discussion of covariance in Appendix A of][]{NG9}. Note that we have fixed the decay timescale and fits to this parameter will further decrease the equivalent $\mathcal{T}(f)$. Since PSR~J1713+0747 has been one of our most sensitive pulsars, the correlation of residuals between pulsars implies that the sensitivity of our PTA is significantly and negatively impacted by any extra model fits, and finding a separate method to correct the TOAs is ideal.

The primary goal of this work is to model the pulse-profile evolution through the April 2021 event and generate corrected TOAs that maintain phase-connected timing despite significant changes in pulse morphology.
In Section~\ref{Gaussian Model}, we describe our Gaussian component model and the framework used to handle the evolving pulse profile.
Section~\ref{observations} summarizes the timing observations of PSR~J1713+0747. 
In Section~\ref{timing}, we present our procedure to calculate absolute TOAs and produce timing solutions across the profile-shape change event and evaluate its performance relative to standard template-matching techniques. 
Finally, Section~\ref{discussion}, we discuss potential methodological improvements, the implications of our results and relevance of these findings for future PTA datasets. In Appendix~\ref{sec:appendix}, we derive approximate systematic timing biases from mismodeled template shapes and argue that the bias in our method introduces small errors relative to the stochastic fitting uncertainties.

\section{Gaussian Component Fit Modeling} \label{Gaussian Model}

To quantify the structure and evolution of the profile of PSR~J1713+0747, we model the pulsar emission using a Gaussian component framework. 
This approach represents the pulse profile as a sum of individual Gaussian components, offering a convenient way to parameterize the profile structure.
This representation is not intended to correspond to specific physical emission regions, but instead serves as a flexible phenomenological description of the pulse morphology. 
Tracking the amplitudes, centroids, and widths of these components provides a compact and interpretable model of the pulse shape. 
The evolution of these parameters can then be examined as a function of observing frequency and time, forming the basis for the analysis described in this section.

\subsection{Template Variations Across Frequency and Time}

Pulse profiles evolve as a function of observing frequency due to both intrinsic emission physics and propagation effects through the ISM \citep{2010arXiv1010.3785C}. 
Accurate modeling of the pulse profile therefore requires accounting for variations across both observing frequency and time. 
In traditional precision pulsar timing analyses, some frequency-dependent profile variations are corrected for in the timing model.
Frequency-dependent parameters empirically absorb unmodeled chromatic delays by fitting polynomial functions of observing frequency \citep{NANOGrav:2020gpb, NANOGrav:2023hde}. 
Some timing analyses use frequency-resolved narrowband templates, with independent templates constructed for individual frequency channels \citep{EPTA:2023fyk}. 
However, wideband timing methods model the pulse profile as a continuous function of observing frequency, allowing frequency-dependent profile evolution to be incorporated directly into the template \citep{Pennucci:2014dja, Agazie:2025yem}.
In contrast to the time-invariant template methods discussed above, our goal is to model pulse-profile evolution directly as a function of both observing frequency and time through Gaussian component fitting. 
By describing the evolving pulse shape in the profile domain prior to TOA generation, we aim to reduce the need for additional timing-model parameters that can absorb power from the gravitational-wave signals of interest.

To model these effects, we distinguish between template profiles and observed profiles.
Wideband portrait templates are frequency-dependent, noise-free representations of the pulse profile, and serve as the reference for Gaussian component fitting.
These templates describe the intrinsic structure of the pulse shape across the full receiver bandwidth.
Templates are represented as two-dimensional functions of rotational phase and observing frequency, $\nu$ \citep{2014MNRAS.443.3752L, Pennucci:2014dja, Pennucci:2018zow}. 

Figure~\ref{fig:toas} shows that the post-event timing perturbations evolve gradually with time, supporting the assumption that the underlying pulse shape also evolves gradually. 
We model the temporal evolution of the profile through Gaussian component fits to the observed profiles at each epoch. 
This approach allows the wideband templates to describe the frequency-dependent structure of the profile, while Gaussian component fits characterize the temporal evolution of the pulse morphology.

\subsection{Pulse Profile Model} \label{profilemodel}

Average pulsar profiles are well approximated by a superposition of Gaussian-shaped components \citep{Krishnamohan, Kramer_1, 1998AJ....116.1984W, Nathan:2023qwn}.
We model the profile as a sum of $N_{c}$ Gaussian components, formalized in Equation~\ref{gceqn}.
The profile intensity as a function of pulse phase, $\phi$, observing frequency, $\nu$, and time, $t$, is expressed as:
\begin{eqnarray} \label{gceqn} 
 & & \!\!\!\!\!\! I(\phi|\nu,t) \approx \nonumber \\
  & & \!\!\!\!\!\! \sum_{i=1}^{N_{c}} A_{i}(\nu,t) \exp\left(-\frac{1}{2}\left[\frac{\phi_{i}(\nu,t) - \mu_{i}(\nu,t)}{\sigma_{i}(\nu,t)}\right]^{2}\right)\!\!,
\end{eqnarray}
where each component $i$ is described by three parameters: amplitude $A_{i}$, center $\mu_{i}$, and width $\sigma_{i}$, which are functions of observing frequency $\nu$ and time $t$.

We adopt $N_{c}=3$, which we show adequately models the pulse shape while maintaining stable and sufficiently precise timing performance (Figures~\ref{fig:temp_lband} and \ref{fig:temp_800}; see Appendix~\ref{sec:appendix} for a heuristic argument approximating the systematic bias due to mismodeling the true pulse template with a finite number of Gaussian components).

Gaussian component modeling is most stable when the pulse profile is centered in phase. 
Centering the profile avoids boundary effects, reduces degeneracies between the parameters and eliminates wrap-around artifacts.
To prepare the pulse profiles for Gaussian component fitting, we normalize both the templates and observed profiles to unit peak amplitude and rotate the profiles using an integer-bin shift that places maximum intensity at pulse phase 0.5. 
This shift is used only for alignment in the component-fitting procedure and is removed when generating TOAs.
Profiles are not explicitly aligned using dispersion measure prior to fitting, allowing for frequency-dependent delays to be accounted for when estimating TOAs, this is further discussed in Section \ref{timing}. 

\begin{figure}
    \centering
    \includegraphics[width=1.0\linewidth]{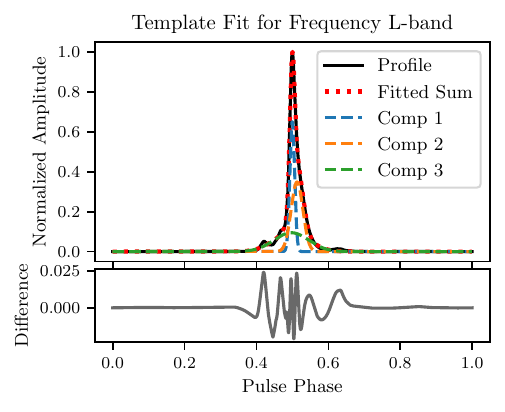}
    \caption{\textit{Top Panel:} Example of a wideband portrait template for L-Band (in black). The Gaussian fit components 1, 2, and 3 are shown by the blue, orange and green dashed lines, respectively. The sum of the Gaussian components is shown by the red dotted line. \textit{Bottom Panel:} The difference between the template (solid black) and fitted sum (dotted red) of the Gaussian components for the L-band shown in the top panel.}
    \label{fig:temp_lband}
\end{figure}

\begin{figure}
    \centering
    \includegraphics[width=1.0\linewidth]{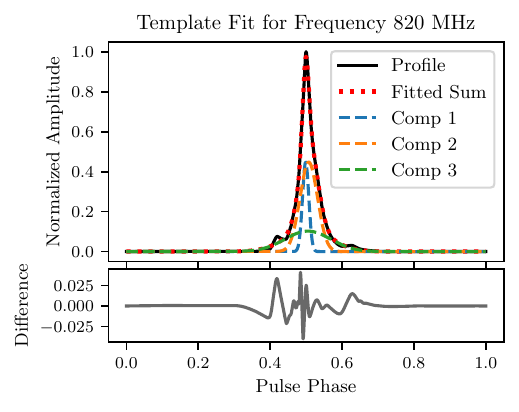}
    \caption{\textit{Top Panel:} Example of a wideband portrait template for the 820~MHz (in black). The Gaussian fit components 1, 2, and 3 are shown by the blue, orange and green dashed lines, respectively. The sum of the Gaussian components is shown by the red dotted line. \textit{Bottom Panel:} The difference between the template (solid black) and fitted sum (dotted red) of the Gaussian components for the 820~MHz band shown in the top panel.}
    \label{fig:temp_800}
\end{figure}

To maintain a consistent ordering of components across all epochs and frequencies, we impose the constraints $A_{1} > A_{2} > A_{3}$ and $A_{i} > 0$ for all $i$.
These constraints prevent component swapping in phase and ensure positive amplitudes throughout the fit.
Larger values of $N_{c}$ can introduce shifts in the centroid of the dominant component, often resulting in increased timing errors. 
Larger values also become more difficult to reliably fit for low signal-to-noise (S/N) pulses.

In the three-component Gaussian model, components~1 and 2 describe narrower, sharply peaked parts of the pulse profile, while component~3 captures a broader, lower amplitude underlying structure. 
The decomposition is intended to provide a compact description of the evolving pulse morphology for timing analysis rather than a unique physical interpretation of the emission geometry.

\section{Observations of PSR J1713+0747} \label{observations}

The data used for the following analysis are part of the 20-year NANOGrav dataset and the observations utilized were made using the Green Bank Telescope (GBT).
We use the Green Bank Ultimate Pulsar Processing Instrument \citep[GUPPI;][]{2008SPIE.7019E..1DD} pre-event wideband portraits as the reference templates for Gaussian component fitting, since equivalent Versatile GBT Astronomical Spectrometer \citep[VEGAS;][]{6051280, 7303578} portrait templates are not currently available. We  assume that the intrinsic pulse-profile morphology remains consistent between the GUPPI and VEGAS backends.

Observations of PSR~J1713+0747 were conducted with the VEGAS using two receiver systems: the PF1 prime focus receiver in the 820~MHz band and the L-band Gregorian focus receiver \citep{Jennings:2024stq}. 
The 820-MHz receiver has a center frequency of 820~MHz and a total bandwidth of 200~MHz, while the L-band receiver is centered at 1500~MHz with a bandwidth of 800~MHz. 

The observations analyzed here span between January 12, 2019 and December 29, 2024 (MJD 58495$-$60673), encompassing epochs both before and after the April 2021 profile-shape change event. 
The observation cadence for the L-band receiver was roughly weekly before the event and monthly afterward, while the 820~MHz receiver was observed approximately monthly over the entire period.
Only data acquired with the GBT and with the VEGAS backend were included in this work; observations from other telescopes or backend systems typically used by NANOGrav were not analyzed. 
Additional details regarding previous observations, coherent dedispersion, pulse calibration, radio-frequency interference mitigation, and data reduction procedures are described in \cite{NANOGrav:2023hde}.

Since our goal is to model the pulse profile evolution through the event and generate corrected TOAs across the shape change, we restrict the analysis to GBT observations obtained with the VEGAS backend, which provides coverage both before and after the event. 
This restriction ensures a homogeneous instrumental configuration across the dataset and minimizes systematic effects from backend-dependent profile shapes and calibration differences. 
Timing offsets for earlier datasets can still be derived using standard timing methods, since the pulse profile remained stable prior to the event and therefore does not require explicit modeling of profile evolutions.

\section{Timing Solutions} \label{timing}

Traditional high precision-pulsar timing is typically performed by measuring the shift of the observed profile relative to a fixed, high S/N averaged template \citep{1983ApJS...53..169D, NANOGrav:2023hde}.
These templates are invariant in time and frequency, and are noiseless. 
Differences between the observed and predicted TOAs are interpreted as arising from, e.g., rotational, astrometric, orbital, or propagation effects, among others, described by the timing model \citep{NANOGrav:2017wvv, NANOGrav:2020gpb}. 

Template-matching procedures assume that the pulse profile is stable. 
During the April 2021 event, the profile of PSR~J1713+0747 changed measurably over time, violating this assumption. 
In the case of an evolving profile, the standard timing method can produce biased TOAs, as changes in the pulse morphology can mimic apparent shifts in arrival time. 
In this regime, timing residuals can no longer disentangle changes in the pulse arrival time from true rotational or propagation-induced variations.

To address this limitation, we adopt a profile-domain timing approach in which pulse shape variations are modeled explicitly using Gaussian components.
Instead of reducing each observation into a single TOA, we decompose the pulse profile into a small number of Gaussian components and track these components relative to frequency-dependent templates. 
Phase offsets derived this way provide a direct measurement of profile evolution, which are then converted into TOAs for standard timing analysis.

\subsection{Reference Template Fitting and Initial Conditions}

In order to provide reasonable initial conditions for the Gaussian component fitting routines, we first fit the noise-free GUPPI L-band standard template. 
The pulse profile is also shifted such that the maximum amplitude occurs at 0.5 phase. 
The iterative fitting routine utilized by \texttt{PyPulse} \citep{pypulse} returns best-fit parameters for $N_{c}$ Gaussian components, where we fix $N_{c}=3$.
To test the robustness and explore a wider portion of parameter space, we employed a Markov chain Monte Carlo (MCMC) fit with the constraints $A_{1} > A_{2} > A_{3}$.
For reference templates the MCMC converges to the same solution obtained by \texttt{PyPulse}, thus validating the stability of the fit.

The resulting parameter estimates, along with their stochastic uncertainties, are listed in Table~\ref{tab:gc_inital}. 
These uncertainties reflect the precision of the MCMC fit but do not include contributions associated with potential systematic errors with the model not matching the template. 
The parameter estimates in Table~\ref{tab:gc_inital} serve as the initial conditions for the Gaussian component fits applied to both the L-band and 820~MHz wideband portraits (Figures~\ref{fig:temp_lband} and \ref{fig:temp_800}), as well as the corresponding observational profiles.

\begin{table}[h]
\centering
\begin{tabular}{c c c c}
\hline
$i$ & $A_i$ & $\mu_i$ & $\sigma_i$ \\
\hline
1 & 0.6482(2)   & 0.499605(1) & 0.006590(1) \\
2 & 0.3495(1)   & 0.512802(6) & 0.016285(5) \\
3 & 0.09588(8)  & 0.49691(1)  & 0.04848(2)  \\
\hline
\end{tabular}
    \caption{Table of initial conditions for the parameters of the Gaussian component fitting model determined by MCMC for L-band.}
    \label{tab:gc_inital}
\end{table}

\subsubsection{Template Fitting} \label{tempfit}

For each receiver band, wideband portrait templates are independently fit using the Gaussian component modeling method described in Section~\ref{profilemodel}. 
For each receiver, the templates are divided into 64 frequency channels to capture the frequency-dependent evolution of the pulse profile while remaining time independent. 
Across the 64 frequency channels, we extract the best-fit amplitudes, centroids, and widths for each of the three Gaussian components, along with their uncertainties which are derived from the covariance matrix.
Figures~\ref{fig:temp_lband} and \ref{fig:temp_800} illustrate the fit results of the center frequency template for L-band (1100--1900~MHz) and 820~MHz (720--920~MHz) receivers, respectively. 
The bottom panels show the residuals between the template and the sum of the fitted components, highlighting the level of agreement between the model and template profile. 

\subsubsection{Observational Data Fitting}

The same fitting routine is applied to the observational data. 
Due to the presence of noise from various sources, additional quality cuts are implemented to ensure robust fits. 
Profiles with a S/N ratio (defined as the peak-to-off-pulse RMS) below 7 are excluded from the analysis, as Gaussian component fits become increasingly unstable at lower S/N and produce poorly constrained centroid and width estimates.
Similar S/N thresholding procedures are commonly used in conventional PTA timing analyses to remove low-quality TOAs \citep{NANOGrav:2020qll}
The adopted threshold balances retaining sufficient data coverage while minimizing poorly constrained fits that can introduce large timing uncertainties. 

\subsection{Relative Residuals} \label{rel_resids}

To study the evolution of the pulse profile through the April 2021 event, we first analyze relative phase residuals rather than relying directly on residuals from a global timing model, following a similar methodology presented in \cite{Lam:2017duo}. 
This technique allows the structure of the phase perturbations to be examined as a function of both observing frequency and time while minimizing biases introduced by an incomplete timing model. 
In particular, conventional timing models are not designed to capture transient pulse-profile evolution and may absorb portions of the signal into other fitted timing parameters. 
We therefore assume that the existing timing solution provides sufficiently aligned folded and dedispersed pulse profiles and focus on relative changes between profiles to isolate variations associated with the pulse-shape evolution. 

For each observation, the centroid of each Gaussian component is measured for both the wideband portrait template and the observed profile.
For a given component $i$, frequency $\nu$, and time $t$, the relative phase residual is defined as:
\begin{equation} \label{phase offset}
    \Delta \phi_{i}(\nu, t) = \mu_{i, \mathrm{obs}}(\nu, t) - \mu_{i, \mathrm{temp}}(\nu) 
\end{equation}
where $\mu_{i, \mathrm{obs}}$ is the observed centroid, and $\mu_{i, \mathrm{temp}}$ is the corresponding template centroid.
Throughout this work, phase residuals are expressed in units of pulse rotational phase (rot), where a value of one corresponds to one full pulsar rotation.
The uncertainty on the relative phase is computed by adding (i) the Gaussian centroid uncertainties determined from the least-squares fit and due to finite S/N in quadrature, with (ii) systematic errors that arise because the Gaussian model does not perfectly reproduce the pulse profile:
\begin{equation}
    \sigma_{\Delta \phi_{i}(\nu,t)} = \sqrt{\sigma_{\mu,i,\mathrm{S/N}(\nu, t)}^2 + \sigma_{\mu,i,\mathrm{sys}(\nu, t)}^2}.
\end{equation}
The relative phase residual therefore quantifies the phase shift required to align a Gaussian component in the observed profile with the corresponding component in the template profile.
Positive and negative residual indicate that a component appears later or earlier in the rotational phase relative to its template position, respectively. 
The uncertainty in Equation~\ref{phase offset} reflects the statistical precision with which the Gaussian centroids are measured in the observed and template profiles. 
There is an additional systematic error that arises because the Gaussian model is fit to the template shape as well and does not perfectly reproduce the pulse profile; however, the resulting template offset does not change in time and so the bias is absorbed into $\mu_{i,\mathrm{temp}}(\nu)$. 

At low S/N, the lowest-amplitude Gaussian component ($i=3$) becomes increasingly difficult to constrain reliably and exhibits larger phase residual variations than the other components.
To systematically remove unstable fits, we apply a data quality cut based on the behavior of this component.
Specifically, for any given epoch and frequency channel, if the absolute phase residual of component~3 exceeds $|\Delta \phi_{3}| > 0.025$, it is excluded from further analysis along with the corresponding component~1 and 2 residuals.
Applying this cut ensures that only well-constrained profiles contribute to the analysis, improving the reliability of the relative phase residual measurements.

\begin{figure}
    \centering
    \includegraphics[width=1.0\linewidth]{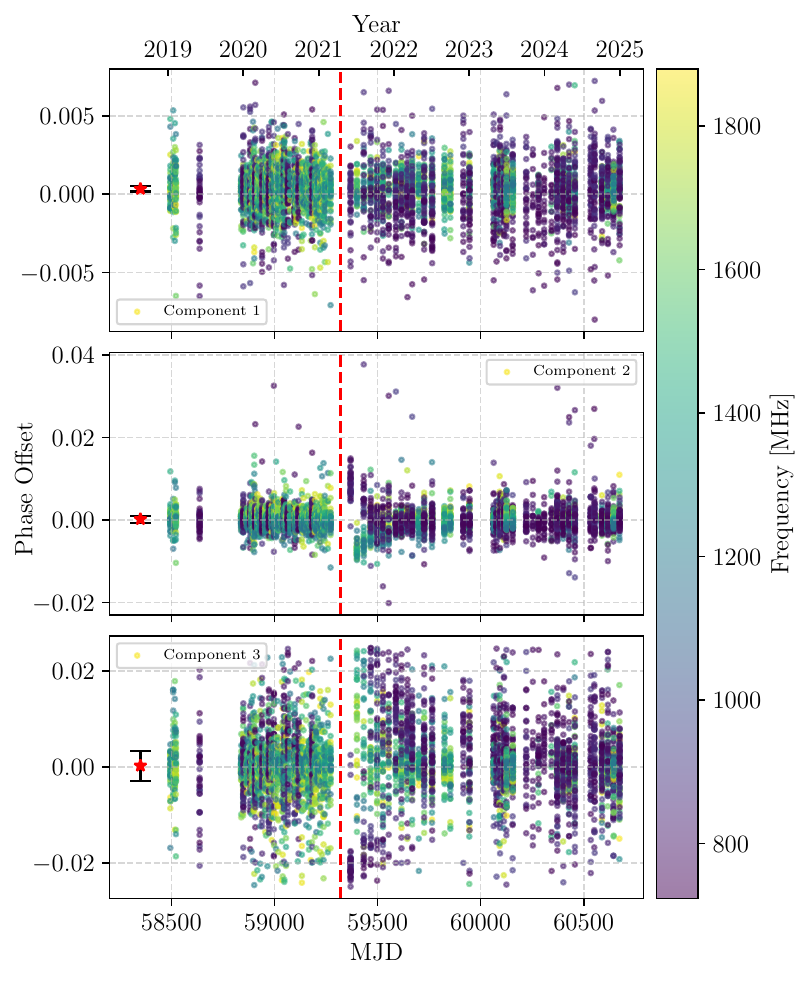}
    \caption{Relative residuals for each Gaussian component derived from our Gaussian component fitting method. The color bar denotes the frequency of residual, with lower frequencies in purple and higher frequencies in yellow. The date of the event is given by the vertical dashed red line. The median phase offset (red point) and uncertainty (black error bars) are plotted prior to the start MJD of the data for each component. Note that the y-axis scales are different for each component.}
    \label{fig:relative_resids}
\end{figure}

Figure~\ref{fig:relative_resids} shows the relative residuals along with the median residual and uncertainty for each component. 
Component~1 appears to exhibit the most stable phase evolution across the event, maintaining continuous phase connection throughout the dataset. 
The median phase residuals for components~1, 2, and~3 are $\sim 3.4 \times 10^{-4}$, $\sim 1.9 \times 10^{-4}$, and $\sim 2.8 \times 10^{-4}$ rot, respectively. 
Although component~2 has the smallest median residual, component~1 has the smallest uncertainty on the median, $\sim 1.6 \times 10^{-4}$, compared to $\sim 8.7 \times 10^{-4}$ and $\sim 3.1 \times 10^{-3}$ rot for components~2 and~3. 
These larger uncertainties, together with the increased scatter in their phase residuals, make components~2 and~3 less suitable for precision timing through the event.

In conventional timing analyses, formal TOA uncertainty estimates begin by quantifying the template-fitting contribution to the TOA uncertainty under idealized conditions, while the scatter observed in timing residuals reflects additional white-noise processes not captured by template fitting alone.
This discrepancy motivates the inclusion of extra white-noise terms in the timing model \citep{NANOGrav:2017wvv, NANOGrav:2023ctt}.
In PTA analyses, these additional white-noise contributions are commonly described using the EFAC, EQUAD and ECORR parameters, which provide a phenomenological model for capturing deviations from the idealized TOA uncertainty model.
EFAC is a multiplicative scaling factor that accounts for the underestimation of template-matching errors due to low S/N observations and discrepancies due to profile variability.
For uncertainties not captured by EFAC, EQUAD represents an additional white-noise contribution which is added in quadrature to the TOA uncertainties. 
Finally, ECORR models noise that is correlated among TOAs obtained within the same observing epoch. 
The white-noise covariance matrix is thus given by \footnote{Some literature definitions apply EFAC only to the formal TOA uncertainty and not to EQUAD.}
\begin{equation}
    C_{ij} = \mathcal{F}_{\mathrm{rcvr}}^{2}\left[\sigma_{\Delta\phi_{i}(\nu, t)}^{2} + \mathcal{Q}_{\mathrm{rcvr}}^{2}\right]\delta_{ij} + \mathcal{J}_{\mathrm{rcvr}}^{2}  \mathcal{U}_{ij},
\end{equation}
where $C_{i,j}$ is the covariance between TOAs $i$ and $j$, $\sigma_{\Delta\phi_{i}(\nu, t)}$ is the Gaussian-component-derived uncertainty of TOA $i$, and $\delta_{i,j}$ is the Kronecker delta. 
The parameters $\mathcal{F}_{\mathrm{rcvr}}$, $\mathcal{Q}_{\mathrm{rcvr}}$, and $\mathcal{J}_{\mathrm{rcvr}}$ correspond to the EFAC, EQUAD and ECORR noise terms for each receiver, respectively.
The matrix $\mathcal{U}_{i,j}$ is unity when TOAs $i$ and $j$ occur in the same observing epoch and zero otherwise.

\subsection{Modeling Post-Event Behavior} 

The relative phase residuals presented in Section~\ref{rel_resids} demonstrate that the pulse-shape change event introduces potential persistent shifts in the Gaussian component centroids that continue after the initial onset of the event. 
These shifts are most clearly visible in components~2 and 3, where the phase residuals exhibit a clear post-event displacement followed by gradual trend toward a quasi-stable state. 
If left unmodeled, these phase offsets propagate directly into the TOAs and will bias the subsequent timing analysis. 
We therefore seek an empirical model of the post-event evolution that captures the dominant behavior of the phase residuals while remaining robust for precision timing applications.
The magnitude of the perturbation varies across the observing band, therefore any correction must account for the frequency dependence of the post-event profile evolution. 

\begin{figure}
    \centering
    \includegraphics[width=1.0\linewidth]{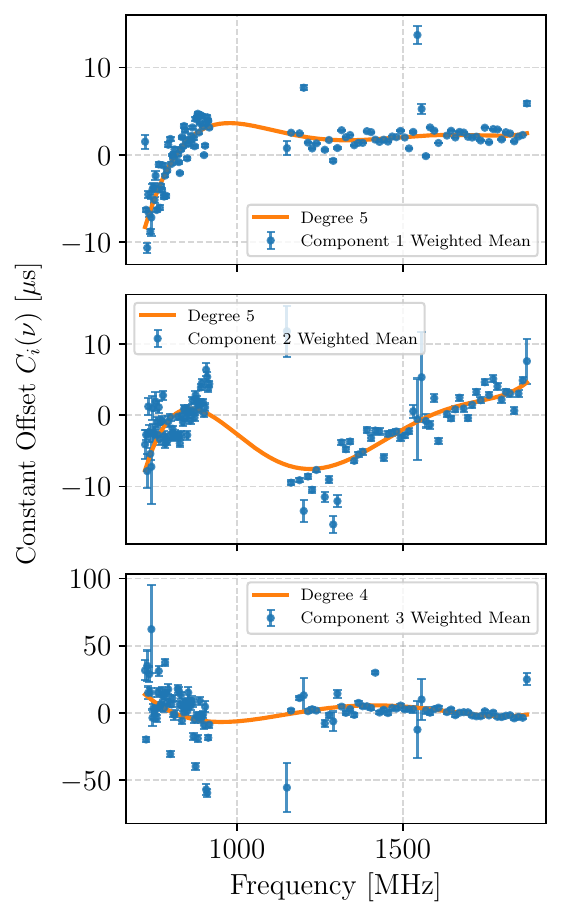}
    \caption{Frequency dependence of the post-event phase offset for all three Gaussian components. Points show the inverse-variance weighted mean phase offset measured in each frequency bin using post-event observations, with error bars indicating the uncertainty on the weighted mean. The solid curve shows the best-fitting polynomial model selected using the AIC. The measured offsets exhibit significant frequency dependence, particularly at lower frequencies, motivating the use of a smooth frequency-dependent correction when generating post-event TOAs.
}
    \label{fig:constfits}
\end{figure}
\subsubsection{Frequency-Dependent Constant Offset Model} \label{constmodel}

Inspection of the post-event phase residuals shows that the most stable feature of the profile evolution is a persistent phase offset whose magnitude varies with observing frequency. 
To characterize this behavior, we calculate the inverse-variance weighted mean phase residual for each Gaussian component and frequency channel using all post-event observations. 
\begin{equation}
    C(\nu) = \frac{\sum_{j=1}^{N_\nu}\Delta\phi_j(\nu,t_j)\,\sigma_{\Delta\phi,j}^{-2}}{\sum_{j=1}^{N_\nu}\sigma_{\Delta\phi,j}^{-2}
},
\end{equation}
where $N_{\nu}$ is the number of post-event observations in frequency channel $\nu$, $\Delta\phi_{j}(\nu,t_{j})$ is the measured phase residual for observation $j$ and $\sigma_{\Delta\phi,j}$ is its uncertainty.
The summation averages over the temporal variability of the post-event residuals, isolating the frequency-dependent phase offset.

The corresponding uncertainty on the weighted mean is 
\begin{equation}                            \sigma_{C(\nu)}=\left(\sum_{j=1}^{N_\nu}\sigma_{\Delta\phi,j}^{-2}\right)^{-1/2}.
\end{equation}

The resulting fits are shown in Figure~\ref{fig:constfits}. 
Significant frequency dependence is present in all three Gaussian components, particularly at lower observing frequencies.
To obtain a smooth correction function, weighted polynomial models were fit as a function of observing frequency.
The polynomial degree was selected using the Akaike information criterion (AIC). 
We emphasize that $C_i(\nu)$ is applied as a correction to the Gaussian component phases prior to TOA generation rather than as an additional parameter in the timing model. 
As a result, the correction is not incorporated into the timing-model design matrix and does not introduce the sensitivity loss associated with fitting additional timing-model parameters as discussed in Section~\ref{intro}. 

We note that the frequency dependence of component~1 in Figure~\ref{fig:constfits} is qualitatively consistent with the pre-event FD behavior reported by \citet{Zhu:2015mdo} though small differences remain between the two curves. However, if no event had occurred, we would expect that $C_1(\nu)$ would be roughly zero for all frequencies (flat) while the FD parameters would still model the general profile evolution. While the FD parameters and $C_i(\nu)$ model the frequency dependence of the profile and components, respectively, the interplay between the two is not immediately obvious. Nonetheless, the amplitude of the curves show demonstrate that the overall $C_1(\nu)$ measures post-event frequency-dependent structure in component~1 comparable to the pre-event profile variations. Given the significance of the offset, we adopt the frequency-dependent constant offset model to generate corrected post-event TOAs for component~1 using a fifth-degree polynomial correction in the subsequent timing analysis.

\subsubsection{Limitations of the Exponential Recovery Model} \label{expmodel}

Previous studies of the April 2021 event reported approximately exponential temporal evolution in the timing perturbations and \citep{Jennings:2022cuj, Jacobson-Bell:2026btw}.
The pulse profile does not appear to return completely to its pre-event morphology. Instead, the phase residuals exhibit a gradual evolution toward a new quasi-stable state whose magnitude varies with observing frequency (Figure~\ref{fig:toas}).
Motivated by these results, we also explored models of the form
\begin{equation}
    \Delta\phi_{i}(\nu, t) = a_i(\nu) \exp\left({-\frac{t-t_{0}}{\tau_{i}}}\right) + C_i(\nu),
\end{equation}
where $a_{i}$, $\tau_{i}$, and $C_{i}$ represent the amplitude, decay timescale and asymptotic phase offset, respectively. 

Weighted exponential models were fit to the recovered parameters as a function of frequency, with the preferred polynomial degree determined again using the AIC. 
The recovered decay timescales have median values of approximately 180, 91, and 117 days for components~1, 2, and 3. 
For components 1 and 3, a frequency-independent timescale (polynomial degree 0) is statistically preferred, while component~2 exhibits modest frequency dependence. 
The amplitudes show weak-to-moderate frequency dependence, with preferred polynomial degrees between 2 and 3. 

In contrast, the constant offset term $C_i(\nu)$ exhibits the strongest and most consistent frequency dependence, with high-order polynomial models preferred for all three components. 
This behavior indicates that the dominant measurable feature of the post-event evolution is a persistent frequency-dependent phase offset rather than a strongly frequency-dependent exponential recovery.
We therefore apply the $C_i(\nu)$ corrections alone, without modeling additional exponential-decay behavior in the timing perturbations.

\subsection{Absolute TOA Generation for Precision Pulsar Timing Use}

The ultimate goal of this analysis is to generate pulse TOAs that can be incorporated into precision pulsar timing datasets for the characterization of the stochastic gravitational wave background. 
While the Gaussian component fitting procedure provides measurements of the relative phase evolution of individual pulse-profile components, pulsar timing requires pulse arrival times reference to an absolute time standard. 
The observational data are stored in PSRFITS files, which record the start time of each subintegration.
To construct TOAs within the Gaussian component timing framework, we therefore associate the measured centroid of component~1 with a pulse arrival time and convert its phase offset into an absolute arrival-time correction. 

When converting the measured phase offsets into absolute pulse arrival times, the rotational evolution of the pulsar must be taken into account. 
As pulsars gradually lose rotational kinetic energy, their spin frequency decreases with time. 
Consequently, the time corresponding to a given phase offset is not constant throughout the dataset. 
To account for this effect, we generate the TOAs directly from the measured phase residuals, rather than comparing against a standard template used in regular fitting procedures, to avoid biases from the evolving profile.  
For each subintegration, the pulse phase model is: 
\begin{equation}
    \phi(t) = f_{0}\Delta t + \frac{1}{2}f_{1}\Delta t^{2},
\end{equation}
where $f_{0}$ and $f_{1}$ are the spin frequency and the spin down rate evaluated at the reference epoch.
We define $\Delta t = t - t_{\mathrm{sub}}$, where $t_{\mathrm{sub}}$ is the subintegration start time referenced to the observation start time.
To recover the timing offset $\Delta t$ corresponding to a measured phase residual $\phi$, we solve the quadratic equation analytically.
In the limit $f_{1} \rightarrow 0$, the solution simplifies to $\Delta t = \frac{\phi}{f_{0}}$. 
However, given the decade timescales considered in the timing model the spin down rate of PSR~J1713+0747 is non-zero.
Thus, we adopt the physically meaningful root,
\begin{equation}
    \Delta t = \frac{2\phi}{f_{0}+\sqrt{{f_{0}^{2}}+2f_{1}\phi}}.
\end{equation}

We generate the TOAs from the measured Gaussian component centroids after removing any phase shifts introduced during the profile-alignment state of the fitting procedure. 
This guarantees that the measured phase residuals reflect the intrinsic component displacement relative to the template. 
To maintain phase continuity, all residuals are wrapped into the interval [0,1).

The absolute TOA is then computed as 
\begin{equation}
    t_{\mathrm{TOA}} = t_{\mathrm{sub}} + \frac{\Delta t}{86400},
\end{equation}
where $t_{\mathrm{sub}}$ is given in MJD. 
TOA uncertainties are propagated from the phase uncertainties assuming $\sigma_{\Delta t} \approx \frac{\sigma_{\phi}}{f_{0}}$ and converted to units of days. 
To minimize numerical errors when adding small phase offsets to large MJD values, all intermediate timing calculations are performed using arbitrary-precision arithmetic (\texttt{Decimal}).
Final component~1 TOAs are rounded to 15 decimal places in MJD, which is the standard for NANOGrav and IPTA format \citep{Antoniadis:2022pcn}. 
The TOAs are written to a \texttt{.tim} file together with their corresponding uncertainties, observing frequencies, and instrumental metadata (frontend and backend).

Timing solutions are generated using pulsar timing packages from \texttt{PINT} \citep{2018AAS...23145309L}.
We adapt the NANOGrav 15-year timing model for PSR~J1713+0747 as the starting ephemeris \citep{NANOGrav:2023hde}. 
The timing model includes frequency-dependent (FD) parameters, which describe the average frequency evolution of the pulse profile and account for systematic chromatic offsets in the TOAs \citep{2015ApJ...813...65N}.
Since the FD parameters were derived using the pre-event NANOGrav 15-year dataset, they are held fixed in our analysis, preserving the established description of the pre-event profile evolution.
Although our dataset includes additional pre-event observations beyond the timespan of the NANOGrav 15-year release, we assume that the underlying frequency evolution the pulse profile remains unchanged prior to the April 2021 event. 
We therefore do not refit the FD parameters, allowing the Gaussian component analysis to isolate the profile-shape perturbations associated with the event.

With the FD parameters now fixed, the remaining timing parameters are fit following standard timing procedures. 
This includes fitting for the pulsar spin, astrometric, and binary parameters, together with DMX parameters that model time-variable dispersion measure variations as piecewise constant offsets \citep{NANOGrav:2023hde, Iraci:2024xhs}.
Since the generated TOAs extend beyond the timespan covered by the NANOGrav 15-year ephemeris, additional per-epoch DMX bins are added and fit for the epochs not included in the original timing model.

A systematic offset is observed between the pre- and post-event TOAs, indicating that the profile-shape change introduces timing perturbations that are not fully described by the initial timing model derived for this analysis.
To account for this behavior, we compare timing solutions containing one jump parameter (between the L-band and 820~MHz datasets) and three jump parameters (between the pre- and post-event datasets in each receiver band).

In the single-jump model, the full L-band dataset is adopted as the reference and a single jump is applied to the 820~MHz TOAs. In the three-jump model, the pre-event L-band TOAs are used as the reference, with independent jumps applied to the pre-event 820~MHz, post-event 820~MHz, and post-event L-band TOAs. 
Either receiver band could be chosen as the reference without affecting the relative timing solution; however, we adopt the L-band as the reference throughout this work. 

The three-jump model is strongly preferred by both the Akaike and Bayesian information criteria (BIC).
Taking the three-jump model as the preferred solution, the single-jump solution has $\Delta\mathrm{AIC}=103.69$ and $\Delta\mathrm{BIC}=90.51$.
This indicates that independent offsets between the receiver bands and the pre- and post-event datasets provide a significantly better description of the TOAs; therefore we use the three-jump model in the final timing analysis (see Table~\ref{tab:jumps}).

\begin{table}[h]
    \centering
    \begin{tabular}{c c}
        \hline
        Receiver Band & Jump Value  \\
        \hline
        820~MHz Pre-event & $-6.0(6) \times 10^{-6}$~s \\
        820~MHz Post-event & $4.0(4) \times 10^{-6}$~s \\
        L-band Post-event & $6.3(2) \times 10^{-6}$~s \\
        \hline
    \end{tabular}
    \caption{Values for the additional jump parameters in the timing model.}
    \label{tab:jumps}
\end{table}

\begin{figure}
    \centering
    \includegraphics[width=1.0\linewidth]{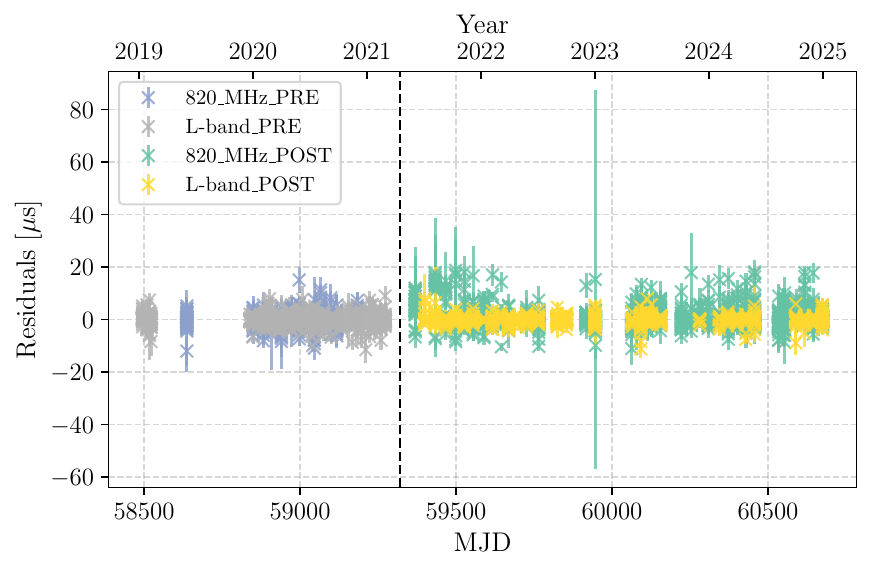}
    \caption{Timing solution for component~1 after refitting the TOAs using a stochastic white and red noise model generated from 20,000 iterations. The fit was initialized from the NANOGrav 15-year timing model with fixed FD parameters, additional DMX bins, and additional jump parameters between pre- and post-event TOAs for both receivers.
}
    \label{fig:postnoise}
\end{figure}

Following the deterministic timing fit, stochastic white- and red-noise processes are modeled using \texttt{enterprise} \citep{2020zndo...4059815E} over 20,000 MCMC iterations. Finally, the timing-model parameters are refit using the resulting noise model (Figure~\ref{fig:postnoise}). 
A small number of catastrophic outliers with absolute post-fit residuals greater than $1000~\mu\mathrm{s}$ were excluded from the residual statistics, removing 15 of 5368 TOAs ($<0.3\%$ of the dataset).

\begin{figure}
    \centering
    \includegraphics[width=1.0\linewidth]{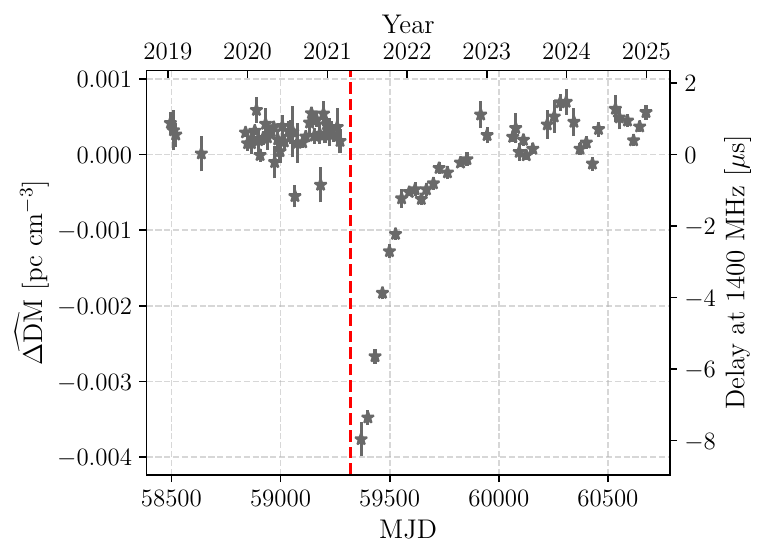}
    \caption{Estimated dispersion measure (DMX) time series for PSR~J1713+0747 obtained by refitting the Gaussian component-derived TOAs with the final post-noise timing model. The vertical red dashed line marks the onset of the profile-shape change event at MJD~59320. A discontinuity is visible near the event epoch. We stress that the estimated DMX parameters are absorbing remaining $\sim\nu^{-2}$ timing delays during the event, and potentially at later times as well; consequently, they should not be interpreted as purely representing true dispersion-measure variations
}
    \label{fig:dmx}
\end{figure}

As an additional diagnostic of the final timing solution, Figure~\ref{fig:dmx} shows the DMX time series obtained by refitting the Gaussian component-derived TOAs using the post-noise timing model. 
A pronounced negative excursion is observed immediately following the profile-shape change event at MJD~59320, followed by a gradual recovery toward the pre-event level. 
The recovered DMX variations exhibit a morphology similar to that reported by \cite{Lam:2017duo} for the previous two chromatic events.
Any remaining shape variations unmodeled by our Gaussian components with a $\sim \nu^{-2}$ dependence will be covariant with DM estimation, similar to the low-level variations shown for the second chromatic event by \citet{Lin:2021vzj}. However, an apparent dip of $4 \times 10^{-3}$~pc~cm$^{-3}$ corresponds to a time delay of 8.5~$\mu$s at 1400~MHz, for example, or approximately 4 phase bins of time delay. 
This amplitude is larger than the expected difference from profile mismodeling (see Appendix~\ref{sec:appendix}) but comparable to the overall time delay shown by the constant offset $C_1(\nu)$. 
We hypothesize that there is significant interplay between these offsets, receiver jumps, and DM modeling. 
We discuss this interplay further in Section~\ref{discussion}.

\subsection{Method Timing Precision} \label{performance}

To assess the suitability of Gaussian component-derived TOAs for precision pulsar timing applications, we incorporate the component~1 TOAs into a full timing analysis and compare their timing precision with that obtained from conventional template matching procedures. 
We perform this comparison using component~1, which remains phase connected throughout the dataset and exhibits less pronounced post-event variability than the other components, making it the most suitable for precision timing analyses.

The post-fit residuals (Figure~\ref{fig:postnoise}) have an RMS in phase of $5.56\times10^{-4}$, which corresponds to $2.54~\mu$s.
The final timing solution remains phase connected across the profile-shape change event and yields a reduced $\chi^{2}$ of 1.02 after noise modeling and refitting.
Although red-noise parameters were included in the noise model, the corresponding posteriors were not constrained or significant, and are not discussed further.
The fitted white-noise parameters have EFAC values of 1.12 -- 1.72 and EQUAD values below $0.54~\mu$s.
The ECORR values range from $0.03~\mu$s to $0.51~\mu$s for the L-band data and from $1.05~\mu$s to $1.93~\mu$s for the 820~MHz data.
Together, these results indicate that the adopted noise model adequately describes the residual scatter. 

\begin{figure}
    \centering
    \includegraphics[width=1.0\linewidth]{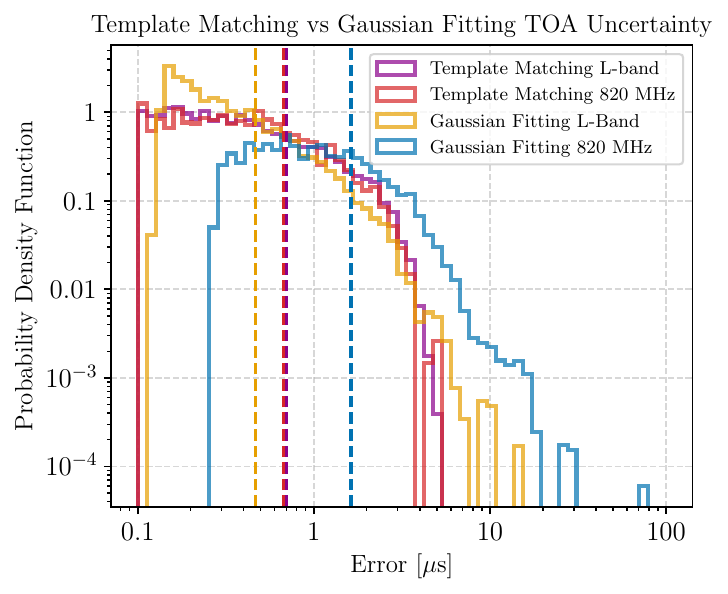}
    \caption{Probability density distributions of the TOA uncertainties obtained from template matching for the L-band (purple histogram) and 820~MHz (red histogram), and from the Gaussian component timing method for the L-band (gold histogram) and 820~MHz (blue histogram). The median of each distribution is shown by the dashed vertical line in the corresponding color. Note that both the x and y axes are logarithmic. 
}
    \label{fig:error_hist}
\end{figure}

As shown in Figure~\ref{fig:error_hist}, the Gaussian component timing method produces TOA uncertainties comparable to those obtained with standard template matching at L-band, with a median uncertainty of $\sim $0.47~$\mu$s compared to $\sim $0.70~$\mu$s for template matching. 
At 820~MHz, however, the Gaussian component-derived TOA uncertainties are larger, with a median of $\sim $1.63~$\mu$s compared to $\sim $0.67~$\mu$s for template matching. 
This increase is consistent with the stronger profile evolution and broader pulse morphology at lower frequencies. 
Overall, these results demonstrate that Gaussian component-derived TOAs can produce stable timing solutions suitable for precision pulsar timing while mitigating the effects of profile variation.

\section{Discussion and Conclusions} \label{discussion}

Pulse-profile evolution in time presents an increasing challenge for high-precision PTA timing analyses as timing baselines extend and measurement precision improves \citep{Brook:2018aho, Jacobson-Bell:2026btw}. 
Methods that preserve phase connection while minimizing systematic timing offsets during periods of irregular behavior are crucial for maximizing the available data for analyses. 
The Gaussian component timing approach explored in this work provides a practical framework for addressing pulse-profile variability without discarding valuable observational data. 
In the following subsections, we discuss both prospects for further methodological refinement and the astrophysical implications of this method for future PTA datasets. 

\subsection{Prospects for Methodological Improvements}

Abrupt pulse-profile variations challenge the assumption of a stable pulse template used in conventional timing analyses.
In this work, component~1 provides the most stable timing solution across the event and maintains phase continuity despite significant changes in the overall pulse morphology. 
As discussed in Section~\ref{performance}, the precision of Gaussian component timing is comparable to template matching for L-band observations, but degrades for the 820~MHz data.
This component-level method could potentially be refined to further improve the timing precision.

Shown in Figures~\ref{fig:temp_lband}~and~\ref{fig:temp_800}, there is a small but measurable difference between the observed pulse profile and the sum of the fitted Gaussian components. 
Increasing the number of components in the fit improves the accuracy of the Gaussian model template, however this comes at the expense of reduced timing sensitivity for individual components and reduced stability of the Gaussian decomposition.
Using the iterative fitting method implemented in \texttt{PyPulse}, the residuals between the GUPPI L-band standard template and Gaussian model fall below 0.01 in normalized amplitude across all pulse phases for $N_c = 7$. 
Further enhancements may be achieved by implementing timing with a weighted combination of components, reflecting that each component contributes unequally to the pulse TOA, instead of only timing with one component.
While increasing the total number of fitted components allows for finer pulse-profile structure to be modeled, the resulting improvement in template fidelity must be balanced against the increased complexity of the Gaussian decomposition and the reduced stability of individual model parameters. 

In Section~\ref{timing}, we discuss the process of determining the initial conditions for the Gaussian component model. 
The GUPPI L-band standard template is used as the reference for both the L-band and 820~MHz data. 
In Figure~\ref{fig:postnoise}, a group of positively shifted post-event 820~MHz TOAs is visible, which may be related to the more pronounced low-frequency profile evolution on the trailing edge of the pulse (Figure~\ref{fig:profiles}). 
The median TOA uncertainty at 820~MHz is also approximately 2.5 times larger than at L-band (Section~\ref{performance}). 
A potential improvement would be to construct receiver-specific Gaussian component templates, allowing the 820~MHz data to be modeled using an independent low-frequency reference profile. 
Since the profile evolution is more pronounced at lower frequencies, a dedicated 820~MHz template may reduce the systematic residual structure and improve the corresponding TOA uncertainties.

The methodology presented here treats the frequency- and time-dependent profile evolution separate from the timing-model offsets (JUMPs) and apparent DMX variations in the analysis. 
The requirement for epoch- and receiver-dependent jump parameters, the introduction of a post-event frequency-dependent constant offset model, and the recovered DMX excursion all occur at the $\mu$s level, suggesting that these effects may be connected through residual mismodeling of the evolving pulse profile. 
This is perhaps unsurprising given that there is significant covariance between frequency-dependent effects -- profile evolution modeled either by FD parameters or wideband portraits, timing jumps, and DM variations (e.g., \citealt{2025PASA...42..108R}; \citealt{2026ApJ..1002....2A}; S.~V.~Sosa Fiscella et al. in prep.) -- even for standard timing analyses. 
Improving the profile-domain description of the pulse-shape evolution may reduce the need for empirical timing corrections and help prevent some of the profile evolution from being absorbed by chromatic timing parameters such as DMX, but either a joint profile and timing method or a profile-domain timing method \citep{2017MNRAS.466.3706L} may be required to reach even higher timing accuracy for this pulsar.

We demonstrate our component-level modeling using only data from the VEGAS backend of the GBT.
Gaussian component timing determines offsets relative to the pre-event VEGAS templates, while conventional timing methods determine offsets between instrumental backends. 
This allows for a hybrid timing framework where component-level timing is applied during epochs affected by the profile evolution and traditional template matching is used elsewhere in the observing baseline. 
Such an approach could be extended to multi-telescope PTA datasets if pulse-shape variations are observed. 
Incorporating observations of profile variations from multiple telescopes can improve the timing solution by increasing the number of TOAs, providing better temporal coverage, and reducing sensitivity to telescope-specific instrumental systematics.
Multi-telescope datasets also provide broader frequency coverage, which can improve dispersion measure modeling and help distinguish between propagation effects and intrinsic profile evolution. 
It is crucial to ensure that component alignment across telescopes is consistent to account for potential frequency-dependent evolution of individual components, as misalignment could introduce additional systematic errors. 
Altogether, these approaches provide several avenues for improving Gaussian component timing techniques in future PTA analyses. 

\subsection{Astrophysical Implications for Future PTA Datasets}

PSR~J1713+0747 is one of the predominant pulsars in current PTA datasets and consistently provides high timing precision \citep{Zhu:2015mdo}. 
As a result, PSR~J1713+0747 contributes disproportionately to PTA sensitivity to the nanohertz gravitational-wave background, even with two previous lower-level chromatic events. 
Its importance makes the treatment of anomalous or transient behavior particularly consequential for future analyses. 

The April 2021 pulse-profile variation in PSR~J1713+0747 demonstrates how transient profile evolution can affect high-precision PTA timing analyses if not properly modeled, as shown in Figure~\ref{fig:toas}.
Even partial mitigation, for example to the level of the previous events, allows us to retain future PSR~J1713+0747 data at comparable sensitivity to pre-event data. In general, incorporating observations affected by profile evolution into PTA datasets remains important for preserving long timing baselines and maximizing sensitivity to nanohertz gravitational waves (Figure~\ref{fig:transmission}).

More broadly, this work emphasizes the importance of flexible timing methodologies for future PTA datasets.
As timing baselines grow and sensitivity improves, rare but impactful profile shape-change events may become increasingly common. 
Approaches that remain robust to pulse-profile variability will play an increasingly important role in maintaining the long-term stability and sensitivity of future PTA datasets. 

\begin{acknowledgments}
We thank Tyler-Símonne Bowman for helpful discussions and feedback that improved the clarity and presentation of this manuscript.
We acknowledge the use of \texttt{ChatGPT-5.5} \citep{ChatGPT} in code optimization and manuscript review. 
We acknowledge support of the NANOGrav Physics Frontier Center through NSF award numbers 2020265 and 2607948. The Green Bank Observatory and the National Radio Astronomy Observatory are facilities of the NSF operated under cooperative agreement by Associated Universities, Inc.

\input{acksJ1713eventtiming_arXiv}
\end{acknowledgments}

\begin{contribution}
S.A.N preformed the analysis, interpreted the results and wrote the manuscript. 
M.T.L. supervised the project, contributed to the analysis, interpretation, and reviewed and edited the manuscript.
I.H.S. reviewed and edited the manuscript.
J.S.H. contributed to discussion of the sensitivity impact analysis. 
All other authors along with S.A.N. created the curated NANOGrav 20-year dataset.
\end{contribution}

\software{\texttt{PyPulse} \citep{pypulse}, \texttt{PINT} \citep{2018AAS...23145309L}, \texttt{hasasia} \citep{Hazboun2019Hasasia}, \texttt{enterprise} \citep{2020zndo...4059815E}, \texttt{ChatGPT-5.5.} \citep{ChatGPT}}

\appendix

\section{Estimating the TOA Uncertainty Bias Between Gaussian Component Modeling and the True Template Shape} \label{sec:appendix}

The Gaussian component fit to the PSR~J1713+0747 pre-event template does not perfectly match its true shape and the resulting perturbations lead to a TOA offset. Since the underlying post-event template shape evolves over time in a way that we cannot perfectly model, the resulting TOA offsets lead to per-epoch systematic biases that should inflate our reported TOA uncertainties. In principle, using more components should better match a true template shape in the regime where S/N approaches infinity, but we are never in that regime on any given epoch. We wish to estimate the biases resulting from using an imperfect Gaussian-component template on the data. Below we derive an order-of-magnitude estimate of this bias.

Our pulsar signal model is
\be
I(t) = a\left[U(t-t_0) +\delta U(t-t_0)\right] + n(t),
\ee
where $I(t)$ is the observed intensity profile, $U(t)$ is the true template shape, $\delta U(t)$ is the template shape perturbation, and $n(t)$ is Gaussian white noise. In the matched filter approach, we assume the template shape is scaled by a factor $a$ and shifted by an amount $t_0$ to match the data shape plus noise, but here we have the extra shape perturbation; we arrive at the usual signal model when $\delta U = 0$. We can extend the template-fitting formalism in the correlation domain \citep[see e.g.,][]{1983ApJS...53..169D,2021ApJ...915...15J} to include a $\delta U$ component, or follow a Fourier domain formalism \citep[see e.g.,][]{1992RSPTA.341..117T,DemorestThesis} but either way we are still left with a numeric calculation to perform and we still do not know how to separate the true template shape from the perturbation.

Instead, consider the approach where we derive TOAs from the function centroids, defined as 
\be 
\left< x \right>_f = \frac{\displaystyle{\int} x\,f(x)\,dx}{\displaystyle{\int} f(x)\,dx}
\ee
for function $f(x)$ following the notation of \citet{2008ApJ...674L..37H}. Since we care about deriving the systematic bias, we ignore the noise $n(t)$ which contributes to the stochastic uncertainty and in practice is not a continuous function. The centroid of the unperturbed pulse shape $I_0(t) = aU(t-t_0)$ is
\be 
\left< t \right>_{I_0} = \frac{\displaystyle{\int} t\,I_0(t)\,dt}{\displaystyle{\int} I_0(t)\,dt}= \frac{\displaystyle{\int} t\,U(t-t_0)\,dt}{\displaystyle{\int} U(t-t_0)\,dt}.
\ee
Now consider our perturbed shape $I(t) = a\left[U(t-t_0) + \delta U(t-t_0)\right] \equiv I_0(t) + \delta I(t)$, where $\delta U(t)$ is small compared to $U(t)$. Then we have that
\be 
\left< t \right>_I = \frac{\displaystyle{\int} t\left[U(t-t_0)+\delta U(t-t_0)\right]\,dt}{\displaystyle{\int} U(t-t_0)+\delta U(t-t_0)\,dt} = \frac{\displaystyle{\int} t\,U(t-t_0)\,dt + \displaystyle{\int} t\delta U(t-t_0)\,dt}{\displaystyle{\int} U(t-t_0)\,dt + \displaystyle{\int}\delta U(t-t_0)\,dt}.
\ee
The second term of the denominator is small by assumption, and the second term of the numerator is small if we assume the pulse perturbation is close to or contained within the same phase range as the main pulse (as is the case for PSR~J1713+0747, $\delta U(t)$ contains intensity at similar phases to $U(t)$). Assuming both to invoke the binomial approximation and ignoring second-order terms, then the centroid of $I(t)$ is
\be 
\left< t \right>_I \approx \left< t \right>_{I_0} + \left(\left< t \right>_{\delta I} - \left< t \right>_{I_0} \right) \cdot \frac{\displaystyle{\int} \delta U(t) \,dt}{\displaystyle{\int} U(t) \,dt},
\ee
noting that the integrals at the end of the expression which give the ratio of the areas are equal whether the function arguments are $t$ or $t-t_0$. The measured centroid (the TOA by our definitional approach) approximately equals the true pulse centroid plus the correction term: the centroid of the pulse perturbation referenced to the true TOA and weighted by the ratio of the areas. 

Applying this equation to the profile residuals for PSR~J1713+0747 (see the bottom panels of Figures~\ref{fig:temp_lband} and \ref{fig:temp_800}), consider the mismatch at the trailing edge of the pulse around phase 0.63, approximately 1/8th of a pulse period away from the center (we have set the pulse peak to be at phase 0.5, and the full pulse centroid is around phase 0.505). The area of $U(t)$ is $\approx 1.0 \times 0.05 = 0.05$ units given the normalization to unit height. The area of $\delta U(t)$ is $\approx 0.01 \times 0.025 = 2.5 \times 10^{-4}$ units so that the ratio of the latter to the former is $5 \times 10^{-3}$. Multiplying by the centroid difference and cast into time units by multiplying by the pulse period, the TOA perturbation from mismodeling this feature is approximately 3~$\mu$s. The other profile residuals are closer to the full pulse centroid and with similar or smaller areas and therefore have a smaller impact on the timing perturbation. Moreover, many of these timing perturbations will cancel with one another, and so we take this as a representative {\it conservative} estimate of the systematic uncertainty using three Gaussian components might produce in our fitting, comparable with the uncertainties shown in Figure~\ref{fig:error_hist}.

\bibliography{sample701}{}
\bibliographystyle{aasjournalv7}

\end{document}

%% file: authorsJ1713event_arXiv.tex
\author[0009-0001-1750-3531]{Shania A. Nichols}
\altaffiliation{NANOGrav Physics Frontiers Center Postdoctoral Fellow}
\affiliation{SETI Institute, 339 N Bernardo Ave Suite 200, Mountain View, CA 94043, USA}
\email{}
\author[0000-0003-0721-651X]{Michael T. Lam}
\affiliation{SETI Institute, 339 N Bernardo Ave Suite 200, Mountain View, CA 94043, USA}
\email{}
\author[0000-0001-5134-3925]{Gabriella Agazie}
\affiliation{Center for Gravitation, Cosmology and Astrophysics, Department of Physics and Astronomy, University of Wisconsin-Milwaukee,\\ P.O. Box 413, Milwaukee, WI 53201, USA}
\email{}
\author[0000-0002-8395-957X]{Anjana Ashok}
\affiliation{Department of Physics, Oregon State University, Corvallis, OR 97331, USA}
\email{}
\author[0000-0002-4972-1525]{Jeremy G. Baier}
\affiliation{Department of Physics, Oregon State University, Corvallis, OR 97331, USA}
\email{}
\author[0000-0002-6039-692X]{H. Thankful Cromartie}
\affiliation{Department of Physics and Astronomy, Vanderbilt University, 2301 Vanderbilt Place, Nashville, TN 37235, USA}
\email{}
\author[0000-0002-1529-5169]{Kathryn Crowter}
\affiliation{Department of Physics and Astronomy, University of British Columbia, 6224 Agricultural Road, Vancouver, BC V6T 1Z1, Canada}
\email{}
\author[0000-0002-2185-1790]{Megan E. DeCesar}
\affiliation{Department of Physics and Astronomy, George Mason University, Fairfax, VA 22030}
\email{}
\author[0000-0002-6664-965X]{Paul B. Demorest}
\affiliation{National Radio Astronomy Observatory, 1003 Lopezville Rd., Socorro, NM 87801, USA}
\email{}
\author[0000-0002-2554-0674]{Lankeswar Dey}
\affiliation{Institute of Astrophysics, FORTH, GR-71110, Heraklion, Greece}
\email{}
\author[0000-0001-5645-5336]{William Fiore}
\affiliation{Department of Physics and Astronomy, University of British Columbia, 6224 Agricultural Road, Vancouver, BC V6T 1Z1, Canada}
\email{}
\author[0000-0001-8384-5049]{Emmanuel Fonseca}
\affiliation{Department of Physics and Astronomy, West Virginia University, P.O. Box 6315, Morgantown, WV 26506, USA}
\affiliation{Center for Gravitational Waves and Cosmology, West Virginia University, Chestnut Ridge Research Building, Morgantown, WV 26505, USA}
\email{}
\author[0000-0003-4090-9780]{Joseph Glaser}
\affiliation{Department of Physics and Astronomy, West Virginia University, P.O. Box 6315, Morgantown, WV 26506, USA}
\affiliation{Center for Gravitational Waves and Cosmology, West Virginia University, Chestnut Ridge Research Building, Morgantown, WV 26505, USA}
\email{}
\author[0000-0003-1884-348X]{Deborah C. Good}
\affiliation{Department of Physics and Astronomy, University of Montana, 32 Campus Drive, Missoula, MT 59812}
\email{}
\author[0000-0003-2742-3321]{Jeffrey S. Hazboun}
\affiliation{Department of Physics, Oregon State University, Corvallis, OR 97331, USA}
\email{}
\author[0000-0003-1082-2342]{Ross J. Jennings}
\altaffiliation{NANOGrav Physics Frontiers Center Postdoctoral Fellow}
\affiliation{Department of Physics and Astronomy, West Virginia University, P.O. Box 6315, Morgantown, WV 26506, USA}
\affiliation{Center for Gravitational Waves and Cosmology, West Virginia University, Chestnut Ridge Research Building, Morgantown, WV 26505, USA}
\email{}
\author[0000-0001-6295-2881]{David L. Kaplan}
\affiliation{Center for Gravitation, Cosmology and Astrophysics, Department of Physics and Astronomy, University of Wisconsin-Milwaukee,\\ P.O. Box 413, Milwaukee, WI 53201, USA}
\email{}
\author[0000-0001-6436-8216]{Bjorn Larsen}
\affiliation{Department of Physics, Yale University, New Haven, CT 06511, USA}
\email{}
\author[0009-0006-9938-157X]{Georgia A. Lowes}
\affiliation{E.A. Milne Centre for Astrophysics, University of Hull, Cottingham Road, Kingston-upon-Hull, HU6 7RX, UK}
\affiliation{Centre of Excellence for Data Science, Artificial Intelligence and Modelling (DAIM), University of Hull, Cottingham Road, Kingston-upon-Hull, HU6 7RX, UK}
\email{}
\author[0000-0001-5229-7430]{Ryan S. Lynch}
\affiliation{Green Bank Observatory, P.O. Box 2, Green Bank, WV 24944, USA}
\email{}
\author[0000-0001-8313-0895]{Ashley Martsen}
\affiliation{Department of Physics and Astronomy, West Virginia University, P.O. Box 6315, Morgantown, WV 26506, USA}
\affiliation{Center for Gravitational Waves and Cosmology, West Virginia University, Chestnut Ridge Research Building, Morgantown, WV 26505, USA}
\email{}
\author[0000-0001-8845-1225]{Bradley W. Meyers}
\affiliation{Australian SKA Regional Centre (AusSRC), Curtin University, Bentley, WA 6102, Australia}
\affiliation{International Centre for Radio Astronomy Research (ICRAR), Curtin University, Bentley, WA 6102, Australia}
\email{}
\author[0000-0002-2689-0190]{Patrick M. Meyers}
\affiliation{ETH Zurich, Institute for Particle Physics and Astrophysics, Wolfgang-Pauli-Strasse 27, 8093 Zurich, Switzerland}
\email{}
\author[0000-0002-0940-6563]{Mason Ng}
\affiliation{Department of Physics, McGill University, 3600  University St., Montreal, QC H3A 2T8, Canada}
\affiliation{Trottier Space Institute at McGill University, 3550 rue University, Montr\'{e}al, QC H3A 2A7, Canada}
\email{}
\author[0000-0002-7374-6925]{Daniel J. Oliver}
\altaffiliation{NANOGrav Physics Frontiers Center Postdoctoral Fellow}
\affiliation{Department of Physics, Oregon State University, Corvallis, OR 97331, USA}
\email{}
\author[0000-0002-8509-5947]{Benetge B. P. Perera}
\affiliation{Arecibo Observatory, HC3 Box 53995, Arecibo, PR 00612, USA}
\email{}
\author[0000-0002-2074-4360]{Henri A. Radovan}
\affiliation{Department of Physics, University of Puerto Rico, Mayag\"{u}ez, PR 00681, USA}
\email{}
\author[0000-0001-5799-9714]{Scott M. Ransom}
\affiliation{National Radio Astronomy Observatory, 520 Edgemont Road, Charlottesville, VA 22903, USA}
\email{}
\author[0000-0001-7832-9066]{Alexander Saffer}
\altaffiliation{NANOGrav Physics Frontiers Center Postdoctoral Fellow}
\affiliation{National Radio Astronomy Observatory, 520 Edgemont Road, Charlottesville, VA 22903, USA}
\email{}
\author[0000-0001-9784-8670]{Ingrid H. Stairs}
\affiliation{Department of Physics and Astronomy, University of British Columbia, 6224 Agricultural Road, Vancouver, BC V6T 1Z1, Canada}
\email{}
\author[0009-0001-5938-5000]{Mercedes S. Thompson}
\affiliation{Department of Physics and Astronomy, University of British Columbia, 6224 Agricultural Road, Vancouver, BC V6T 1Z1, Canada}
\email{}
\author[0009-0002-6412-7812]{Amir Tresnjic}
\affiliation{ETH Zurich, Institute for Particle Physics and Astrophysics, Wolfgang-Pauli-Strasse 27, 8093 Zurich, Switzerland}
\email{}
\author[0000-0002-4088-896X]{Joris P. W. Verbiest}
\affiliation{Florida Space Institute, University of Central Florida, 12354 Research Parkway, Orlando, FL 32826, USA}
\email{}
\author[0000-0003-1562-4679]{David Wright}
\affiliation{Department of Physics, Oregon State University, Corvallis, OR 97331, USA}
\email{}

%% file: acksJ1713eventtiming_arXiv.tex
The work of A.A.\ and D.W.\ is partly supported by the George and Hannah Bolinger Memorial Fund in the College of Science at Oregon State University.
A.A.\ gratefully acknowledges the support of Moore Foundation.
Pulsar research at UBC is supported by an NSERC Discovery Grant and by CIFAR.
K.C.\ and M.S.T.\ are supported by a UBC Four Year Fellowship.
D.C.G.\ is supported by NSF Astronomy and Astrophysics Grant (AAG) award \#2406919.
M.T.L.\ received support by an NSF Astronomy and Astrophysics Grant (AAG) award number 2009468 during this work.
H.A.R.\ is supported by NSF Partnerships for Research and Education in Physics (PREP) award No.\ 2216793.
S.M.R.\ and I.H.S.\ are CIFAR Fellows.
J.P.W.V.\ acknowledges support from NSF AccelNet award No.~2114721.